\documentclass[letterpaper,11pt,fleqn]{article}
\pdfoutput=1
\usepackage{jheppub}

\usepackage{dcolumn}

\usepackage{graphicx}
\usepackage{bm,amsmath,amssymb}
\usepackage[mathscr]{eucal}

 \newcommand{\eqn}[1]{Eq.~(\ref{#1})}
\newcommand{\beq}{\begin{equation}}
\newcommand{\eeq}{\end{equation}}

\newcommand{\rmd}{{\rm d}}

\long\def\comment#1{ }

\title{ Anomalous dimensions from rotating open strings in AdS/CFT}
 
\author[a]{Nelson R. F. Braga}
\author[b]{and Edmond Iancu}

\affiliation[a]{Instituto de F\'{\i}sica,
Universidade Federal do Rio de Janeiro, Caixa Postal 68528, RJ
21941-972 -- Brazil}
\affiliation[b]{Institut de Physique Th\'{e}orique de Saclay,
F-91191 Gif-sur-Yvette, France}

\emailAdd{braga@if.ufrj.br}
\emailAdd{edmond.iancu@cea.fr}
 
\abstract{ 
We propose a new entry within the dictionary of the AdS/CFT duality at strong coupling:
in the limit of a large spin or a large R-charge, the anomalous dimension
of the gauge theory operator dual to a semiclassical rotating string is proportional to the string proper length. 
This conjecture is motivated by a generalization to strings of the rule for computing anomalous dimensions of 
massive particles and supergravity fields in the anti-de Sitter space.
We show that this proportionality holds for a rotating closed string in global AdS space, representing a high spin operator made of fields in the adjoint representation.
It is also valid for closed strings rotating in $S^5$ (representing operators with large R-charge), 
for closed strings with multiple AdS spin, and for giant magnons.
Based on this conjecture, we calculate the anomalous dimension $\delta$ of
operators made of fields in the fundamental representation, associated with high spin mesons, 
and which are represented by rotating open strings attached to probe D7-branes.
The result is a logarithmic dependence upon the spin,  $\delta\sim \sqrt{\lambda}\ln S$,
similar to the closed string case.
We show that the operator properties --- anomalous dimension and spin --- are obtained from measurements 
made by a local observer in the anti-de Sitter space.  For the open string case, this ensures that 
these quantities are independent of the mass scale introduced by the D7-branes (the quark mass), 
as expected on physical grounds. In contrast, properties of the gauge theory states, 
like the energy, correspond to measurements by a gauge theory observer and depend upon the mass 
scale --- once again, as expected. 
  }


\begin{document}
\maketitle

\section{ Introduction }
Anomalous dimensions of high spin operators are important ingredients of gauge theories. 
In perturbative QCD, the anomalous dimension grows with the logarithm of the spin. This property played an essential role in the analysis of the approach to Bjorken scaling
in deep inelastic scattering, for $\vert q^2 \vert \to \infty \,$, as a consequence of asymptotic freedom  \cite{Gross:1974cs,Georgi:1951sr}.
The logarithm dependence for high spins appears to be valid at all loop  orders
\cite{Korchemsky:1988si,Korchemsky:1992xv,Basso:2006nk}. 

At strong coupling, it is possible to investigate gauge theories using the gauge/string duality \cite{Maldacena:1997re,Witten:1998qj,Gubser:1998bc} :
for the case of  ${\cal N} = 4$ supersymmetric Yang Mills (SYM) theory, the AdS/CFT correspondence
provides a `dual' description in terms of a string theory living in the AdS$_5\times$S$^5$ space-time. 
This in particular allows one to compute the 
anomalous dimensions of high-spin operators at strong coupling \cite{Gubser:2002tv}. 
Within the context of the pure--gauge ${\cal N} = 4$ SYM theory, such operators are
exclusively built with fields in the adjoint representation of the gauge group SU$(N)$.
Specifically, Ref. \cite{Gubser:2002tv} computed the anomalous dimensions of high-spin, twist-two, operators 
like $ {\rm Tr} \big( \phi^I D_{(\mu_1 ... }\,D _{\mu_S )}\,\phi^I \big)$ via studies of the corresponding `dual' object 
---  a semiclassical closed string rotating in global  AdS$_5$ space-time. 
Within this duality, time translations in global AdS$_5$ correspond to dilatations in the gauge theory. Accordingly, 
the conformal dimension $\Delta$ of the gauge-theory operator can be identified with 
the energy of the rotating string. In the high-spin limit $S\gg \sqrt{\lambda}$, this method yields a 
logarithmic dependence of the anomalous dimension $\delta=\Delta-S$ upon the spin $S$, which is
similar to that found in perturbation theory (in both QCD and ${\cal N} = 4$ SYM) :
\begin{equation}
\label{anomalousGKP}
\delta \,=\, \Delta - S \, \simeq \,   \frac{\sqrt{\lambda}}{ \pi}\,  \ln \bigg( \frac{S}{\sqrt{\lambda}} \bigg)\,,
\end{equation}
where $\lambda = g_{_{\rm YM}}^2 N$ is the 't Hooft coupling and $g_{_{\rm YM}}$ is the coupling of
the gauge theory.

Within perturbative QCD, a similar behavior at large $S$ holds also for the twist-two operators 
built with quark fields in the {\em fundamental} representation of SU$(N)$, with the general structure
\begin{equation}
\label{OS}
{\cal O}_S \,=\, \bar \Psi  \, \gamma^{(\mu_1} D^{\mu_2} ...D^{\mu_S)} \Psi \,\,,
\end{equation} 
where $\Psi $ denotes the quark field. Once again, one expects this logarithmic
behavior $\delta\propto\ln S$ for $S\to\infty$ to hold to all orders in perturbation theory, and also
at strong coupling (at least, within gauge theories involving matter fields in the fundamental
representation and for which the strong coupling limit is clear to make sense, 
like the ${\cal N} = 2$ Yang Mills theory that we shall focus on in this work). 
Besides explicit calculations in perturbation theory at low orders, 
there are solid non-perturbative arguments in that sense.

One such an argument comes from the relation between the anomalous dimension 
of twist-two operators with high spin $S$ and the anomalous dimension of a cusp in a 
light-like Wilson line \cite{Korchemsky:1988si,Korchemsky:1992xv} :
the coefficient $ f(\lambda ) \,$ in the asymptotic expression
$ \delta \simeq  f(\lambda ) \ln S$ for $\delta$ at large $S$ coincides with the cusp anomalous dimension.
This relation is conjectured to hold regardless of the value of the coupling and for any
gauge-group representation of the matter fields within the structure of the operator. 
(This has been checked in perturbation theory in many gauge theories, 
including non-supersymmetric ones, and also at strong coupling in the context of ${\cal N} = 4$ SYM 
\cite{Drukker:1999zq,Kruczenski:2002fb,Makeenko:2002qe,Belitsky:2003ys,Kruczenski:2007cy}.)
If extrapolated at strong coupling, it predicts that the anomalous dimension 
of the quark-antiquark operator in \eqn{OS} at large $S$ should be
half\footnote{Indeed, within the planar limit $N\to\infty$
under consideration, a Wilson loop in the adjoint representation is equal to the square of the
same loop in the fundamental representation; hence, $ f_{\rm adj}(\lambda )=2
f_{\rm fdt}(\lambda )$.} of the corresponding result for an
adjoint operator, as shown in \eqn{anomalousGKP}.  

Another compelling argument comes from the geometrical interpretation, proposed in \cite{Alday:2007mf}, 
for the $\ln S$--dependence of the anomalous dimension $\delta$ at large $S$, and also for
its relation to the cusp anomaly.  Once again,
this interpretation is expected to hold for any value of the coupling and irrespective of the gauge-group
representation.

Notwithstanding, it would be interesting to
have a direct calculation at strong coupling of the anomalous dimensions associated with high-spin
operators involving fundamental matter fields, cf. \eqn{OS}.
In this paper, we shall propose a method in that sense,
based on the study of rotating {\em open} strings in the AdS$_5\times$S$^5$ 
space-time represented in Poincar\'e coordinates.

To describe our method, let us first recall that, in order to include fields in the fundamental representation of
SU$(N)$ (`quarks') in AdS/CFT, one must introduce probe D7-branes in AdS$_5$ \cite{Karch:2002sh}. The dual gauge theory is the ${\cal N} = 2$ Yang Mills theory with massive quarks. Mesons of low spin are described as fluctuations of the D7-branes while high spin mesons are represented by semiclassical rotating open strings attached to the D7-branes in a Poincar\'e AdS$_5$ space \cite{Kruczenski:2003be} (see also \cite{Erdmenger:2007cm} for a review). The presence of the D7-branes introduces a mass scale in the duality, the quark mass $M_q$, corresponding to the position of the branes along the `radial' direction of AdS$_5$. 
As a consequence,  a rotating open string representing a high spin meson 
is not invariant under dilatations. This situation contrasts with the closed string case, where the dual gauge theory is conformal.  So, the approach pioneered in \cite {Gubser:2002tv},
which consists in computing the anomalous dimension $\delta$ from the dilatation charge $\Delta = S + \delta$  
(which in turn is dual to the energy of a rotating closed string in global AdS$_5$), 
cannot be carried over to the case of an open string.  

As an alternative approach, valid to parametric accuracy in the limit of a large spin and/or large R-charge, we propose a method to compute the anomalous dimension $\delta$ of the gauge theory operator dual to a
semiclassical (closed or open) string from the string proper length $\ell$ in the temporal gauge. 
More precisely, introducing the dimensionless `scale parameter' $\gamma \equiv RT_0\ell$, where $R$ is the 
AdS$_5$ radius and $T_0=1/(2 \pi \alpha^\prime)$ is the string tension, we conjecture that
$\gamma$ and $\delta$ are proportional to each other in the semiclassical limit alluded to above.
The proportionality constant (a pure number) is however not uniquely determined, as we shall see on specific examples.

Our proposal, to be detailed in Sect.~\ref{sec:gamma}, is motivated by the analogy 
with the corresponding calculations for (very) massive supergravity objects which are 
localized in AdS$_5$, like supergravity fields  \cite{Witten:1998qj,Gubser:1998bc}
and point-like AdS particles \cite{Janik:2010gc}.
In these well-known examples, the anomalous dimensions of the respective dual operators
in the semiclassical limit $mR\gg 1$ are obtained as $\delta\simeq mR$, with $m$ the
mass of the supergravity dual. As we shall explain in Sect.~\ref{sec:gamma}, our 
scale parameter $\gamma$ is proportional to the integral over the string of the masses of 
infinitesimal string bits measured by local observers in AdS (so, in that sense, it plays the same
role as the mass of an AdS particle). 
Then, in Sect.~\ref{sec:tests}, we shall test our conjecture on several examples involving 
closed strings (dual to operators made of adjoint fields), 
for which the respective anomalous dimensions have already been computed, 
via the recipe introduced in Ref.~\cite{Gubser:2002tv}. We shall thus demonstrate that
the string length is indeed proportional to the known anomalous dimensions, 
in several cases: a closed string with a large spin $S$  \cite{Gubser:2002tv},
one with a large R-charge \cite{Gubser:2002tv}, 
one with large multiple--spin \cite{Frolov:2003qc},
and finally a giant magnon  \cite{Hofman:2006xt} .

In Sect.~\ref{sec:open},  we address the main physics problem which has motivated our analysis:
the calculation of the anomalous dimension for the quark-antiquark twist-two operator in \eqn{OS} 
at large $S$ and strong coupling. To that aim, we shall assume that the string-theory object dual to 
this operator is an open string rotating in AdS$_5$, 
with the string endpoints attached to a D7-brane. This identification is by no means obvious (in fact,
we are not aware of similar observations in the literature), so let us briefly explain 
our motivation for it. The main argument in that sense is an {\em a posteriori} one,
namely,  the consistency between this (conjectured) duality and our final result for  
$\gamma$ (the scale parameter for a rotating open string). Specifically,
 in the high-spin limit $S\gg \sqrt{\lambda}$,  
we shall find $\gamma\simeq 2({\sqrt{\lambda}}/{ \pi})  \ln \big({S}/{\sqrt{\lambda}} \big)$, 
which is the expected result for the anomalous dimension of the quark operator in \eqn{OS},
to parametric accuracy. The numerical prefactor is probably wrong : as compared
to the respective result for an adjoint operator,  cf. \eqn{anomalousGKP}, we have
a factor of 2, instead of the expected 1/2. But as already mentioned, this numerical coefficient 
is not under control within our method, and should not be trusted.

Another strong argument in favor of the duality between the rotating open string and
the quark-antiquark operator  \eqref{OS} is the fact that the respective scale parameter $\gamma$
is found to be independent of the infrared scale (the quark mass $M_q$) of the gauge
theory, as expected for the anomalous dimension of a {\em local} operator.
This scale invariance, which is manifest on the expression for $\gamma$ in terms of $S$ 
shown above, emerges from the string calculation in a very interesting way:
the proper length $\ell$ of the open string remains invariant under the AdS$_5$
isometry corresponding to dilatations in the gauge theory. The effect of this isometry
is to rescale the radial position of the D7-brane, which in turn corresponds to a redefinition 
of the quark mass $M_q$ in the gauge theory. This invariance is quite 
non-trivial\footnote{This invariance has also been exploited in Ref.~\cite{Kruczenski:2003be},
at least implicitly (via a judicious choice of coordinates), but its consequences 
--- like the scale invariance of the string calculation for the spin ---
have not been discussed there.}
--- it does not hold e.g. for the energy of the string (which scales with the quark mass 
like $M_q$), nor for the separation between the string endpoints on the D7-brane
(which scales  like $1/M_q$ and is indicative of the physical size of the 
meson state in the dual gauge theory). It is related to the fact that, as alluded to above,
the scale parameter $\gamma$ is obtained from string properties as measured by a
{\em local} observer in the AdS space-time. That is, this invariance is tributary to
 the warp factor relating local densities on the string to the corresponding densities 
measured by a gauge-theory observer.

This discussion sheds new light on the way how the AdS/CFT correspondence works in 
the presence of mass scales, as introduced by D7-branes.
A local measurement  in AdS yields results which 
are independent of the mass scale and hence can be attributed to a local operator.
A  gauge--theory observer, on the other hand, measures properties of the gauge-theory state,
like its energy or average size, which are sensible to the mass scale $M_q$.
The case of the spin is particularly interesting in that respect: 
this is at the same time an attribute of the operator, cf. \eqn{OS}, and of the meson state created 
by this operator. And indeed, as we shall further discuss in Sect.~\ref{sec:open}, both local observers
in AdS$_5$, and boundary observers using the gauge-theory time, find the same value $S$ for
the angular momentum of the string.


\section{From the string length to anomalous dimensions}
 \label{sec:gamma}
 
We consider a classical string embedded in the target space-time AdS$_5\times$S$^5$ with metric
$g_{mn}$. For definiteness, in this section we consider  AdS$_5$
with Minkowskian signature and Poincar\'e coordinates, but our discussion below
would also apply to an Euclidean space--time, and also to  AdS$_5$ with global metric. 
A convenient choice of coordinates is
\begin{equation}
\label{AdSmetric}
\rmd s^2 \,=\, \frac{R^2}{z^2} \, ( -\rmd t^2 + \rmd {\bm x}^2 + \rmd z  ^2) + R^2\rmd \Omega_5  ^2
\,\equiv\,g_{mn}\rmd x^m\rmd x^n\,,
\end{equation} 
where $z$ is the holographic radial coordinate, $x^\mu=(x^0,x^1,x^2,x^3)\equiv (t, {\bm x})$ are the
coordinates in the Minkowski space at the boundary of AdS$_5$ at $z=0$, and 
$\rmd \Omega_5^2$ refers to S$^5$ (see \eqn{S5metric} below for an explicit map on S$^5$). 
The string embedding is parametrized as $X^m=X^m (\tau, \sigma) $,
where $\tau$ and $\sigma$ are coordinates on the string world-sheet.
The dynamics of the string is governed by the Nambu-Goto action
\begin{equation}\label{NG}
I =\, -\frac{1}{2 \pi \alpha^\prime}
 \int \rmd \tau \rmd  \sigma \,\sqrt{-\mbox{det}\,
 h_{\alpha\beta}}\,,\qquad
 h_{\alpha\beta}\,=\,g_{mn}\partial_\alpha X^m
 \partial_\beta X^n\,,
\end{equation}
where $\alpha,\,\beta=0$ or 1, $(\sigma^0, \sigma^1)\equiv(\tau,\sigma)$, and
$h_{\alpha\beta}$ is the induced world-sheet metric. Writing $h\equiv \mbox{det}\,
 h_{\alpha\beta}$, one has
\beq\label{h}
-h\,=\,(\dot X\cdot X')^2 - (\dot X)^2 (X')^2\,,\eeq
where a dot (prime) refers to the derivative w.r.t. $\tau$ ($\sigma$),
$A\cdot B\equiv g_{mn}A^m B^n$, and $A^2\equiv A\cdot A$.

We chose the temporal gauge
$\tau = X^0$ and define
\begin{equation}
\label{DimString1}
\gamma \, \equiv \, \frac{R}{2 \pi \alpha^\prime} \,\, \ell \,\equiv\,
\frac{R}{2\pi \alpha' } \,  \int_{_{string}} \rmd \sigma  \, \,\sqrt{ g_{mn}  \frac{ \partial
X^{m }}{\partial\sigma} \frac{\partial X^{n }}{\partial\sigma} } \,.
\end{equation}
For definiteness, the quantities $\gamma$ and $ \ell $ will be referred to as
the `string scale parameter' and, respectively, the `string proper length'.   For what follows, it is 
useful to notice that the above integral can be equivalently rewritten as
\begin{equation}
\label{DimString}
\gamma \,=  \,\,R\, \int_{_{string}} \rmd \sigma  \sqrt{ - g^{mn} \pi^0_m \pi^0_n }\,,
\end{equation}
where $\pi^0_m$ is the density of the $m$--component of the canonical 
momentum on the string world-sheet:
\begin{equation}
\label{densities}
\pi^0_m  \equiv \frac{ \partial  \cal L}{ \partial \dot X^m}\,=\,- T_0 
\,g_{mn}\,\frac{
(\dot X\cdot X')(X^n)' - (X')^2(\dot X^n) }{\sqrt{-h}}
\,,
\end{equation} 
with $\cal L$ the Lagrangian density in \eqn{NG} and $T_0= {1}/{2 \pi \alpha^\prime}$ the
string tension. Indeed, by using Eqs.~\eqref{h} and \eqref{densities}, one can easily check the
identity $\sqrt{ - g^{mn} \pi^0_m \pi^0_n }=\sqrt{(X')^2}$.

For a string living in a {\em flat} space-time with Minkowski metric
$\eta^{mn}$, the equivalence between
the integral expressions in Eqs.~\eqref{DimString1} and respectively \eqref{DimString} 
has a limpid physical interpretation. The quantity $\rmd m\equiv T_0\rmd \ell$ represents the 
mass of an infinitesimal piece of string with length $\rmd \ell=\sqrt{(X')^2}\,\rmd \sigma$. 
Similarly, $\rmd P_m\equiv \pi^0_m \rmd \sigma$ is the (infinitesimal) space-time momentum 
of that piece of string. Hence the relation $\rmd m=\sqrt{-\eta^{mn}\rmd P_m \rmd P_n}$ is recognized
as the mass-shell condition for a relativistic `particle' (here, a bit of the string).
The overall space--time momentum of the string is computed by integrating the momentum density
over the string length: $P_m=\int \rmd\sigma 
\,\pi^0_m$. The mass $M$ of the string is finally obtained as $M=\sqrt{-\eta^{mn}P_m P_n}$.

But for a string living in the curved AdS space-time, the situation is more subtle because
the space--time momenta look different to an inertial observer living in the gauge theory
(`on the Minkowski boundary of AdS$_5$') as opposed to a local observer in ten dimensions.
This difference is due to the gravitational redshift (warp factor) $\sqrt{g_{00}}=R/z$ multiplying the flat
boundary metric in \eqn{AdSmetric}.  Yet, as we now argue, it is natural to interpret 
the integrand in \eqn{DimString} as the {\em local} `mass' 
(or `virtuality') of a bit of the string, as viewed by a local observer in AdS space
(i.e. an observer  whose instantaneous position in ten dimensions
is the same as that of a point on the string and whose local time is $\bar t
\equiv \sqrt{g_{00}} \,t= (R/z)t$). To see this, consider for simplicity a string stretching 
and moving only in AdS$_5$, which has no momenta along the S$^5$ directions~;
for such a string, one has (with sum over  $i=1,2,3$)
\begin{align}\label{dbarm}
 \rmd \sigma  \sqrt{ - g^{mn} \pi^0_m \pi^0_n }\,=\,\rmd \sigma
 \frac{z}{R}\sqrt{(\pi^0_0)^2-(\pi^0_i)^2-(\pi^0_z)^2}\,=\,\sqrt{
({\rmd\bar E})^2-(\rmd{\bar P}^i)^2-(\rmd{\bar P}^z)^2} 
 \,,\end{align}
where
\beq\label{warp}
\frac{\rmd\bar E}{\rmd \sigma}\,=\,-\frac{z}{R}\,\pi^0_0\,,
\quad \frac{\rmd\bar P^i}{\rmd \sigma}\,=\, \frac{z}{R}\,\pi^0_i\,,
\quad \frac{\rmd\bar P^z}{\rmd \sigma}\,=\, \frac{z}{R}\,\pi^0_z
\,,\eeq
are the string densities of the respective canonical momenta,
as seen by the local observer. They differ via the warp factor $z/R$ from the respective
densities which enter the calculation of the {\em physical} 4--momentum, as seen
by a gauge theory observer:
 \beq\label{EP}
E\,=\,-\int \rmd \sigma \,\pi^0_0\,,\qquad P^i\,=\,\int \rmd \sigma \,\pi^0_i\,.
\eeq
This difference is important for our present purposes, in that it shows that the
quantity $ \rmd\bar m\equiv \rmd \sigma  \sqrt{ - g^{mn} \pi^0_m \pi^0_n }$ can
be viewed as the mass of a bit of string in AdS, which in turn suggests an interesting 
physical interpretation for $\gamma=R \int_{_{string}} \rmd \bar m$. 

Namely, it is well known that the AdS masses of {\em local} objects in AdS$_5$, 
like supergravity fields or massive point-like particles, are proportional to the 
anomalous dimensions of the corresponding operators in the dual gauge theory 
\cite{Witten:1998qj,Gubser:1998bc}.
The example of a point particle in AdS$_5$ with mass $m$ is particularly inspiring 
for our present purposes. In the semiclassical limit where $mR\gg 1$, the anomalous dimension
of the CFT operator dual to this AdS particle is found as 
$\delta\simeq mR$ (see \cite{Janik:2010gc} for
an explicit calculation). Note that this mass $m$
is a property measured by a {\em local} observer in AdS$_5$,
in the same way as the infinitesimal string mass $\rmd \bar m$ introduced above.
Indeed, starting with the action for a point particle moving in the $(x^1\equiv x,\,z)$ plane,
\beq I_P = - m \int \rmd \tau \sqrt{ - g_{00} \left( \frac{\partial x^0}{\partial \tau } \right)^2
- g_{xx} \left( \frac{\partial x}{\partial \tau } \right)^2 -
g_{zz} \left( \frac{\partial z}{\partial \tau } \right)^2 } \,\equiv \int \rmd \tau {\cal L}\,,
\eeq
one can easily compute the canonical energy and momenta according to their definitions,
\begin{equation}
\label{Particle_momenta}
P_0 \,= \frac{ \partial  \cal L}{ \partial \dot x^0}
 \ , \quad
P_x = \frac{ \partial  \cal L}{ \partial \dot x} \ , \quad P_z = \frac{  \partial \cal L}{ \partial  \dot z}
 \ ,
\end{equation}
and thus check that the following representation holds for the mass of the particle,
 \begin{equation}
\label{MassParticle}
\sqrt{ - g^{mn} P_m P_n } = \frac{z}{R} \sqrt{  P_0^2  - P_x^2 - P_z^2 } = \,\, m \,.
\end{equation}
Eq.~(\ref{dbarm}) provides a natural generalization of this formula to the case of a string.

In view of the above, we conjecture that the string scale parameter $\gamma$ should have
a similar physical interpretation : for a semiclassical string with large spin, or $R$--charge,
the quantity $\gamma$
should be proportional to the anomalous dimension $\delta$ of the dual gauge theory  operator.  
In the next section, we shall test this conjecture by comparing with known results for single-trace
operators in ${\cal N} = 4$ SYM, as represented by rotating closed strings. 

 
\section{Testing our conjecture versus closed strings}
\label{sec:tests}

In this section, we shall successively consider four different problems involving
closed strings with large angular momenta, which have been studied at length
in the literature, and for which the associated anomalous dimensions have been 
computed via the method introduced in Ref.~\cite{Gubser:2002tv} --- that is,
by exploiting conformal symmetry. For all these strings, we shall compute the 
scale parameter $\gamma$ introduced in the previous section and verify that it
is indeed proportional to the respective anomalous dimensions.

\subsection{Closed string with high spin}  
\label{sec:closed}

For a large spin $S$,
the anomalous dimension of the twist--two operators made of adjoint fields, such as
$ {\rm Tr} \big( \phi^I D_{(\mu_1 ... }\,D _{\mu_S )}\,\phi^I \big)\,$, is known 
to have the form \cite{Benna:2006nd}
\begin{equation}
\label{Cusp}
\Delta \,-S\, =  f(\lambda  ) \, \ln (S) + \mathcal{O}(S^0)\,,
\end{equation}
where $f(\lambda)$ is the cusp anomaly \cite{Korchemsky:1988si,Korchemsky:1992xv}.  
In  the strong coupling limit of ${\cal N} = 4$ SYM theory,  such an operator is dual
to a closed string rotating in global AdS$_5$ space  \cite{Gubser:2002tv}, with metric
\begin{equation}
\label{globalmetric}
\rmd s^2 \,=\,R^2 \, ( -\rmd \tau ^2 \cosh^2 \rho  + \rmd \rho ^2 + \sinh^2 \rho \,\, \rmd \Omega ^2_3\, ) \,.
\end{equation}
The string is folded and stretched thus forming a very thin loop, made of two parallel straight lines, 
which rotates uniformly in the equator plane $( \rho , \phi )$ of  $S^3$ : $ \phi = \omega \tau$.
It exten\rmd s up to a maximum value of the radial coordinate $\rho = \rho_0,\,$ which is related
to the angular velocity via
\begin{equation}
\label{Rel}
\coth^2 \rho_0 \,=\, \omega^2 \,\,.
\end{equation}
It is understood that $\omega > 1$. Note that all the coordinates and also the angular velocity
are dimensionless, and that the physical dimensions are given by appropriate powers of the AdS radius $R$.
Clearly, the string proper length is $\ell=4R\rho_0$,  where the factor of 4 comes by adding the 
four segments of the folded string. This, together with \eqn{DimString1}, implies 
 \begin{equation}
\label{DimClosedString}
\gamma  \,\,=\,\,   4 R \,\int_0^{\rho_0} \rmd \rho    \sqrt{ - g^{mn} \pi^0_m \pi^0_n } = 
\frac{2 R^2 }{ \pi \alpha'}\,\rho_0 \,\,.
\end{equation}
As shown by the first equality above, the same value for $\gamma$ can be inferred from the
string energy--momentum densities that can be read off Eqs.~\eqref{Ecl}--\eqref{Scl} below.

The Nambu-Goto Lagrangian has the form
 \begin{equation}
L \,=\, - \frac{4 R^2}{2 \pi \alpha^\prime} \int_{0}^{\rho_0} \rmd  \rho   \sqrt{ \cosh^2 \rho \,- \omega^2 \sinh^2 \rho  }\,,
\end{equation}
from which the string energy and spin easily follow as (recall \eqn{EP})
\begin{equation}
\label{Ecl}
E \,=\,  -4\int_{0}^{\rho_0}\rmd  \rho  \, \pi^0_0\,=\,
 \frac{4 R^2}{2 \pi \alpha^\prime} \int_{0}^{\rho_0} \rmd  \rho   \,
\frac{\cosh^2\rho }{\sqrt{ \cosh^2 \rho \,- \omega^2 \sinh^2 \rho  } }\,,
\end{equation}
and respectively
\begin{equation}
\label{Scl}
S\,=\,  4\int_{0}^{\rho_0} \rmd  \rho   \,\pi^0_{\phi} \,=\,
 \frac{4 R^2}{2 \pi \alpha^\prime} \int_{0}^{\rho_0} \rmd  \rho  \, 
 \frac{\omega  \sinh^2 \rho }{ \sqrt{ \cosh^2 \rho \,- \omega^2 \sinh^2 \rho }}\,.
\end{equation} 
The anomalous dimension is then obtained as $\delta=E-S$. (We recall that the energy $E$ in global AdS$_5$
coincides with the total dimension $\Delta$ of the dual operator, 
and that $\delta=\Delta-(S+2)\simeq \Delta-S$ for twist-two operators with large spin $S\gg 1$.) 

There are two interesting limits (we recall that $R^2 / \alpha^\prime = \sqrt{\lambda}$) :

\noindent \texttt{(i)} {\em Large string with high spin $S\gg\sqrt{\lambda}$.} This correspon\rmd s to the limit
$\omega\to 1$. Writing $ \omega  = 1 + 2 \eta \,,\,\,$ with $\eta \ll 1 $, we deduce from 
Eq.~(\ref{Rel}) that $\rho_0$ is large,
\begin{equation}
\rho_0 \,= \, \frac{1}{2}\ln \left(
 \frac{1}{ \eta } \right) + {\cal O}(1) \,,
\end{equation}
and from Eqs.~\eqref{Ecl}--\eqref{Scl} that the energy and spin are large as well \cite{Gubser:2002tv} :
\begin{equation}\label{ESclosed}
E \,=\,   \frac{\sqrt{\lambda}}{2 \pi } \left[ \frac{1}{\eta} + \ln \left( \frac{1}{\eta }\right) + ...
\right] \,,\qquad
S \,=\,   \frac{\sqrt{\lambda}}{2 \pi } \left[ \frac{1}{\eta} - \ln \left( \frac{1}{\eta }\right) + ...
\right] \,\,,
\end{equation}
By using these equations,  the scale factor in \eqn{DimClosedString} can be cast into the form
\begin{equation}
\label{DimClosedStringfinal}
\gamma \simeq \,   \frac{ \sqrt{\lambda} }{ \pi }\,\ln \left(
 \frac{S}{  \sqrt{\lambda}} \right) \,.
\end{equation}
This is identical with the respective result for $\delta=E-S$, as emerging from   
\eqn{ESclosed} \cite{Gubser:2002tv}. Note that the way how this matching works
is quite non-trivial: the dominant terms, of ${\cal O}\big(1/\eta\big)$,
in the energy $E$ and the spin $S$ precisely cancel in their difference,
so the anomalous dimension  $\delta$ is sensitive to the {\em subleading} terms of ${\cal O}\big(\ln (1/\eta)\big)$.
By contrast, the scale parameter $\gamma$ is only logarithmically divergent as $\eta\to 0$ and
hence, to the accuracy of interest, it is sensitive to the {\em leading} term in $S$ alone.
What truly matters for our present purposes is that $\gamma$  and $\delta$ are {\em proportional} to
each other, and hence have the same parametric dependencies upon $ \sqrt{\lambda}$ and
${S}$. The fact that the proportionality coefficient appears to be exactly one is
likely to be a coincidence (this will not happen for other examples below). Technically,
this happens because, when
$\omega\to 1$, the integrand in Eqs.~\eqref{Ecl}--\eqref{Scl} for the difference $E-S$ approaches
unity, and then the ensuing integral over $\rho$ is exactly the string length.

\noindent \texttt{(ii)} {\em Small string with relatively low spin $S\ll\sqrt{\lambda}$.} This correspon\rmd s to the 
limit of a large angular velocity $\omega\gg 1$. Then Eq.~(\ref{Rel}) implies $\rho_0\simeq 1/\omega$, 
while Eqs.~\eqref{Ecl}--\eqref{Scl} yield \cite{Gubser:2002tv} 
 \begin{equation}
E \,\simeq\,   \frac{\sqrt{\lambda}}{\omega} \,,\quad
S \,=\,   \frac{\sqrt{\lambda}}{2\omega^2} \,\ll \, E
\,,\eeq
and therefore $\gamma\simeq (2/\pi)\big( {\sqrt{\lambda}}/{\omega}\big)$ is again proportional with
$\delta\simeq E\simeq  {\sqrt{\lambda}}/{\omega}$, but the proportionality factor is now
different from unity. In physical units, one has $\gamma\sim\delta \sim \lambda^{1/4} \sqrt{S}$.
 Notice that this anomalous dimension is still large,  $\delta\gtrsim \lambda^{1/4}$,
albeit not as large as in the case of a high spin, cf. \eqn{DimClosedStringfinal}.
 

\subsection{Closed string with large R-charge }
\label{sec:Rcharge}

As another example, which has also been treated in Ref.  \cite{Gubser:2002tv},
let us consider a closed string rotating in $S^5$ with a large angular momentum $J$ and which is
dual to a CFT operator with a large R-charge, equal to $J$.
Choosing the metric on $S^5$ as
\begin{equation}\label{S5metric}
R^2 \rmd \Omega_5^2 \,=\,R^2 \, 
(\rmd \theta   ^2  + \sin^2 \theta \, \rmd \phi ^2 + \cos^2\theta \, \rmd \Omega ^2_3\, ) \,,
\end{equation}
the string is folded, it rotates uniformly in the $(\theta,\phi)$--plane with $ \phi =  \omega t$,
and it is stretched along the $\theta$ direction up to a value $\theta_0$ given by
\beq
\sin\theta_0\,=\,\frac{1}{\omega}\,.\eeq
The proper length is
$\ell = 4 \int_0^{\theta_0} \sqrt{g_{\theta\theta }}\,\rmd \theta    = 4 R \theta_0$,
where the factor of 4 comes from the four pieces of the folded string. The corresponding
expressions for the string energy $E$, angular momentum $J$, and anomalous dimension
$\delta=E-J$, can be found in Ref.  \cite{Gubser:2002tv}.
As before, there are two interesting regimes, \texttt{(i)} $\omega\to 1$ and
\texttt{(ii)} $\omega\gg 1$, and in both cases the scale parameter $\gamma =({R}/{2 \pi \alpha^\prime})
\ell$ turns out to be proportional to $\delta$. Specifically, in the limit of very large R-charge $J\gg \sqrt{\lambda}$,
 corresponding to $\omega\to 1$, one fin\rmd s $\theta_0\to \pi/2$ and hence
\begin{equation}
\label{DeltaApprox4}
\gamma   \, \simeq \, \frac{R^2}{\alpha'}\,=\,\sqrt{\lambda}\,.
\end{equation} 
This differs by a factor $\pi/2$ from the corresponding result for $\delta=E-J$  \cite{Gubser:2002tv}.
Once again, this agreement at parametric level is quite non-trivial: separately, $E$ and $J$ involve leading-order
contributions which diverge logarithmically in the limit $\omega\to 1$, but which mutually cancel in their
difference.
In the opposite limit where $\omega\gg 1$ and $J\ll E\ll \sqrt{\lambda}$, one fin\rmd s $\theta_0\simeq 1/\omega$, hence
$\gamma\simeq (2/\pi)\big( {\sqrt{\lambda}}/{\omega}\big)$, which differs by a factor $2/\pi$ from the
correct anomalous dimension in this regime, namely $\delta\simeq E \simeq  {\sqrt{\lambda}}/{\omega}$
 \cite{Gubser:2002tv}.

\subsection{Multi-spin closed string }
\label{sec:multi}

Consider now a multi-spin closed string, with two angular momenta $S_1$ and $S_2$ 
corresponding to rotations in two different directions of  AdS$_5$,  as studied in
detail in  Ref. \cite{Frolov:2003qc}. We focus on the string solutions 
describing a circular closed string located at a fixed value of the AdS$_5$ radius, 
while rotating simultaneously in two planes in AdS$_5$ with equal spins $S_1 = S_2\equiv
S$. (This solution is a direct generalization of a two-spin flat-space solution where the string rotates in 
two orthogonal planes while always lying on a 3-sphere.) Since the AdS$_5$ radius is fixed,  $\rho=\rho_0$,
it suffices to consider the $S^3\, $ sector of the global metric  in Eq. (\ref{globalmetric}),
that is
\begin{equation}
\rmd \Omega_3^2\,=\, \rmd \theta   ^2 + \sin^2 \theta \rm\rmd \phi  ^2 + \cos^2 \theta d \varphi^2 \,.
\end{equation} 
The string 
is rotating in the $\phi $ and $\varphi$ angles and winding once around the circle $0 \le \theta < 2\pi $. 
So, the string length is
$\ell =\int_0^{2\pi} \sqrt{g_{\theta\theta}}\,\rmd \theta    =2 \pi  R \sinh \rho_0$.
As announced, the spin components corresponding to rotations in the two angles have both the same magnitude,
\begin{equation}
S_1 = S_2 = \frac{R^2}{\sqrt{2}\alpha'}  \,\sinh^3 \rho_0\,,
\end{equation}
so the total spin  strength is $ S =(R^2/\alpha' ) \,\sinh^3 \rho_0 $. The scale parameter $\gamma$ can be written as
\begin{equation}
 \gamma \,=\,\frac{R}{2\pi \alpha'} \ell \,=\, \frac{R^2}{\alpha'}  \sinh \rho_0 \,=\, 
\frac{R^2}{\alpha'} \left( \frac{S}{R^2/\alpha'} \right)^{1/3},
\end{equation} 
which has the same parametric form as the respective anomalous dimension
computed in Ref. \cite{Frolov:2003qc}, but is larger than the latter by a factor of $2 \sqrt{2}/3 $.

 \subsection{Giant Magnon}
 \label{sec:magnon}
As an additional example of a closed string solution with a large angular momentum in the ten dimensional 
AdS$_5\times$S$^5$ space, let us now consider a giant magnon.
This solution, discussed in detail in  Ref. \cite{Hofman:2006xt},  has an infinite angular momentum in the S$^5$
space, corresponding to an infinite R-charge in the dual gauge theory.  The S$^5$ metric can be written 
in the form of  Eq. (\ref{S5metric}). For the solution possesing the least energy, one can use the results of 
 Ref. \cite{Hofman:2006xt}, and consider a fixed time string profile. From this solution at constant time, 
one finds the following string length: 
\begin{equation}
\ell = 2 \int_{\theta_0}^{\pi/2} \sqrt{ g_{\theta\theta} \rmd \theta   ^2
\, +\, g_{\phi\phi} \rm\rmd \phi  ^2 } =2 R \,\int_{\theta_0}^{\pi/2}
\frac{\sin\theta \,\cos\theta_0}{\sqrt{\sin^2\theta - \sin^2\theta_0}}
\rmd \theta    \, ,
\end{equation}
where the coordinate $\theta$ varies from $\pi/2$ to $\theta
_0 $ and back to  $\pi/2$, and we used the relation between the angles
$\phi$ and $\theta$ for fixed time.
Performing this integral, one fin\rmd s $\ell = R \pi \cos\theta_0$ and therefore
$\gamma = R^2 \cos\theta_0 / 2 \alpha' \,=\,
\sqrt{\lambda} \cos\theta_0 /2 $. This result for the scale factor $\gamma$ is equal to the anomalous
dimension found in \cite{Hofman:2006xt} multiplied by a factor of
$\pi/2$.

\section{Open strings and high spin mesons}
 \label{sec:open}
In this section we shall consider open strings rotating in AdS$_5$,
associated with mesons.  
 
 The system we want to describe consists of open strings rotating in AdS$_5$
with endpoints attached to a probe D7 brane, which are dual to mesons 
 in the ${\cal N} = 2$ Yang Mills theory having a very large space-time spin $S$. 
As stated in the introduction, we expect these dual mesons to correspond to the 
leading-twist operators in the fundamental representation, as shown in \eqn{OS}.
More arguments supporting these interpretation will be provided by our results below.


In order to describe our string configuration, we work with AdS$_5$ space in Poincar\'e coordinates and choose the metric as
\begin{equation}
\label{Pmetric}
\rmd s^2 \,=\, \frac{R^2}{z^2} \, ( -\rmd t^2 + \rmd z  ^2 + \rmd r^2 + r^2 \rmd \theta   ^2 + \rmd y^2 \, )\,,
\end{equation}
\noindent where the spatial coordinates on the Minkowsky boundary are in cylindrical form: ($r,\theta, y$).
A high-spin meson is represented by a semiclassical rotating string with a steady profile 
$( r =r(z),\, \theta = \omega t \,,y=0)$, which is symmetric around $r=0$ and  reaches a maximum value 
$ z = z_0$ in the `radial' direction \cite{Kruczenski:2003be}.   The string endpoints are attached to a D7-brane 
located at $z = z_{D7}$. The quark mass is related to the radial location of the brane via
\begin{equation}
M_q \,=\, \frac{R^2}{2 \pi \alpha^\prime z_{D7} }\,\,.
\end{equation}
As discussed in \cite{Kruczenski:2003be}, it is convenient to use the dimensionless coordinates 
$ \tilde r \equiv \omega r \,,\, \tilde z \equiv \omega z  $. Chosing $\tau= t$ and $\sigma = \tilde z $, the  
Nambu-Goto action takes the form
\begin{equation}
\label{ActionOpen}
I =  -\, \frac{R^2\omega}{ \pi \alpha'}\,\, \int \rmd t \int_{{\tilde z}_{D7}}^{{\tilde z}_0} 
\rmd {\tilde z} \,  \frac{1}{{\tilde z}^2} 
\sqrt{ ( {\tilde r}^{\,\prime \, 2} + 1)( 1 - {\tilde r}^{2})}\,,
\end{equation}
where a prime denotes a derivative with respect to $\tilde z $ and a factor of 2
has been included to account for the string symmetry around $r=0$.
The string angular momentum is computed as
(below $\pi^0_{\theta} =\partial\cal L/\partial\omega$)
\begin{equation}
\label{AngularMomentum}
S = 2\int_{{z}_{D7}}^{{z}_0} 
\rmd {z}    \,\pi^0_{\theta} \,=\,   \frac{R^2 }{ \pi \alpha'}\,\, \int_{{\tilde z}_{D7}}^{{\tilde z}_0} 
\rmd {\tilde z} \,  \frac{{\tilde r}^{2}}{{\tilde z}^2} 
\sqrt{\frac{{\tilde r}^{\,\prime \, 2} + 1}{1 - {\tilde r}^{2}}}\,.
\end{equation}
The string profile $ \tilde r ( \tilde z ) $ is determined by solving the following equation of motion:
\begin{equation}
\label{eqmotion}
\frac{{\tilde r}^{\,\prime\prime}}{1 +  {\tilde r}^{\,\prime \, 2} }
- \frac{2 {\tilde r}^{\,\prime }}{\tilde z} + 
\frac{\tilde r}{1 -  {\tilde r}^{2} }\,=\,0\,,
\end{equation}
with a boundary condition stating that the string ends orthogonally on the D7-brane:
\begin{equation}
\label{BC1}
\frac{\rmd \tilde r }{\rmd \tilde z }\bigg\vert_{_{\tilde z = {\tilde z}_{_{D7}}}} \,\,=\,0\,\,,
\end{equation} 
and the following conditions at $ {\tilde z}_{0}$, which express the symmetry of the string profile:
\begin{eqnarray}
\label{BC2}
\tilde r ( {\tilde z}_0 ) &=& 0 \,\,, \\
\label{BC3}
\frac{\rmd \tilde r }{\rmd \tilde z }\bigg\vert_{_{\tilde z = {\tilde z}_{0}}} &\to& - \infty \,\,.
\end{eqnarray} 
Eqs.~\eqref{eqmotion}--\eqref{BC3} determine the maximal string penetration ${\tilde z}_0$
as a function of the position ${\tilde z}_{_{D7}}$ of the D7-brane, and the string profile $ \tilde r (\tilde z)$
for ${\tilde z}_{_{D7}}\le \tilde z\le {\tilde z}_0$. In particular, the separation $L=
2\tilde r ({\tilde z}_{_{D7}})/\omega$
between the string endpoints on the D7--brane is also interesting, in that it measures the typical size
of the dual meson state in the gauge theory.
We refer to Ref. \cite{Kruczenski:2003be} for detailed numerical studies of the solutions and also for analytical approximations (some of which will be useful in what follows).

Calculating the string densities defined in Eq. (\ref{densities}), one finds the scale parameter $\gamma$ as
\begin{equation}
\label{DimStringOpen}
\gamma \,\equiv \,\, R \,\int_{string}  \rmd \sigma  \sqrt{ - g^{mn} \pi^0_m \pi^0_n } \,\,= \,\, 
\frac{R^2 }{ \pi \alpha'}\,\, \int_{{\tilde z}_{D7}}^{{\tilde z}_0} 
\rmd {\tilde z} \,  \frac{1 }{\tilde z} 
\sqrt{{\tilde r}^{\,\prime \, 2} + 1}\,.
\end{equation}

From now on, we shall focus on operators with large spin  $S\gg\sqrt{\lambda}$. As explained in 
Ref. \cite{Kruczenski:2003be}, this corresponds to the limit of a small angular velocity $\omega\to 0$~;
more precisely,  the relevant, small, parameter is ${\tilde z}_0\ll 1$. This limit can be studied via
a limited expansion in powers of ${\tilde z}_0 $ around the well-known solution ${\tilde r}_{\rm st} (\tilde z )$
corresponding to a static string \cite{Maldacena:1998im,Rey:1998ik}. In this regime, the integral in
\eqn{DimStringOpen} is proportional to the large logarithm $\ln({1}/{{\tilde z}_{0}})$. Indeed, when
${\tilde z}_0\ll 1$, the lower limit ${\tilde z}_{D7}$ is even smaller: one has  
$\tilde z_{_{D7}} \simeq C \, {\tilde z}_0^{\,3} $ with $C $ = 0.599  \cite{Kruczenski:2003be}. Also, to
leading logarithmic accuracy, one can neglect the
derivative ${\tilde r}^{\,\prime \, 2} $ under the square root, in spite of the fact that it diverges
at the endpoint ${\tilde z}_0 $, as required by Eq.~(\ref{BC3}). To see this and also evaluate
\eqn{DimStringOpen} to the accuracy of interest, one can replace ${\tilde r}^\prime $ inside the integrand 
by the respective static solution \cite{Maldacena:1998im,Rey:1998ik}: 
\begin{equation}
 {\tilde r}_{\rm st}^\prime (\tilde z ) = \frac{ {\tilde z }^2 }{\sqrt{{\tilde z }^4_0 - {\tilde z }^4}}\,.
\end{equation}
\noindent (We have checked this approximation against the numerical calculation of the integral in Eq. (\ref{DimStringOpen}).) Then, one can successively rewrite the integral in eq (\ref{DimString}) as follows
 \begin{align}
\label{DeltaApprox1}
\int_{{\tilde z}_{D7}}^{{\tilde z}_0} 
\rmd {\tilde z} \,  \frac{1 }{\tilde z} 
\sqrt{{\tilde r}_{\rm st}^{\,\prime \, 2} + 1} &\,=\frac{1}{2}
\int_\varepsilon^1 \frac{dy}{y}\frac{1}{\sqrt{1 - y^2}}\,\simeq\, 
\frac{1}{2}\int_\varepsilon^1 \frac{dy}{y}
\,+\,\frac{1}{2}\int_0^1 \frac{dy}{y}\Big( \frac{1}{\sqrt{1 - y^2}} - 1 \Big)
\nonumber\\ &\,=\, \ln \bigg( \frac{{\tilde z}_{0}}{{\tilde z}_{D7}}  \bigg)
  \,+\,D\,\,,
\end{align}
where $ y \equiv ({\tilde z}/{\tilde z_0})^2$, $\varepsilon\equiv ({{\tilde z}_{D7}}/{\tilde z}_{0})^2$,
and in the last integral we were allowed to replace the lower limit by zero because $\varepsilon\ll 1$ and 
the integral is regular at that endpoint.
Then this last integral is a pure number, evaluated as  $D \simeq 0.347 \,$. 
We thus obtain
\begin{equation}
\label{DeltaApprox2}
\gamma   \, =\, \frac{ 2\, \sqrt{\lambda}}{ \pi}\, \left[ \ln \bigg( \frac{1}{{\tilde z}_{0}} \bigg) + {\cal O}(1)
\right]\,,
\end{equation} 
where we have also used $\tilde z_{_{D7}} \simeq C \, {\tilde z}_0^{\,3} $ and $\sqrt{\lambda}=
{R^2 }{\alpha'}$. Still in this regime at $ {\tilde z_0 } \ll 1 $, one can estimate the spin
as \cite{Kruczenski:2003be}:
\begin{equation}
\label{JApprox1}
S  \, \simeq \,\frac {\sqrt{\lambda} C}{ \pi {\tilde z}_0} \,\,\,,
\end{equation} 
which confirms that $S\gg\sqrt{\lambda}$, as anticipated. Let us also quote the respective
result for the separation between the string endpoints on the D7-brane: $\tilde r ({\tilde z}_{_{D7}})
\simeq C {\tilde z}_0$  \cite{Maldacena:1998im,Rey:1998ik,Kruczenski:2003be}.

By eliminating ${\tilde z}_0$ between Eqs.~\eqref{DeltaApprox2} and \eqref{JApprox1},
one can express the scale parameter in terms of the spin  $S$~:
\begin{equation}
\label{DeltaApprox3}
\gamma   \, \simeq \, 2 \,\frac{\sqrt{\lambda}}{ \pi}\, \ln \bigg( \frac{S}{\sqrt{\lambda}} \bigg)\,.
\end{equation} 
This rewriting makes it clear that this parameter $\gamma$ is independent of the infrared scale 
in the gauge theory (the quark mass $M_q$), a necessary condition for
the anomalous dimension of a {\em local} operator. It furthermore exhibits the right
parametric dependencies upon $\lambda$ and $S$ (in particular, the characteristic
$\ln S$--behavior) to be identified with the anomalous dimension of the quark-antiquark 
twist-two operator in \eqn{OS}, up to a numerical coefficient.  Hence the result
in \eqn{DeltaApprox3} simultaneously supports
our conjectured dualities between the rotating open string and the local operator \eqref{OS}
and, respectively, between the scale parameter $\gamma$ and the anomalous dimension 
$\delta$. As before, this last identification holds in the sense of a proportionality relation,
in which the numerical coefficient is not under control. And indeed, the overall coefficient in
\eqn{DeltaApprox3} differs by a factor of 4 from the correct respective value, as
expected from the relation with the cusp anomaly for Wilson lines in the fundamental
representation (cf. the discussion in the Introduction).

By inspection of  Eqs.~\eqref{DeltaApprox2}--\eqref{JApprox1}, it is clear that
the dimensionless quantity ${\tilde z}_{0}$ plays the same role within this open-string context
as the (dimensionless) angular velocity $\omega$ within 
the closed--string calculation in Sect.~\ref{sec:closed} : this is a numerical AdS
parameter which \texttt{(i)} specifies the string theory object (here, an open string) which is dual to a
gauge theory operator with a given spin, and \texttt{(ii)} parameterizes
the relation between the spin and the anomalous dimension within the supergravity calculation.
This parameter can be eventually eliminated between $S$ and $\gamma$, and then it totally 
disappears, as it should, from physical predictions such as \eqn{DeltaApprox3}.
From this perspective, ${\tilde z}_{0}$ 
should be viewed as a pure number in AdS, on the same footing as $mR$ in the case of
an AdS particle (or a supergravity field) with mass $m$. 
There is however an interesting difference between these two dimensionless quantities,
$mR$ and respectively ${\tilde z}_{0}=\omega z_0$, which reflects the peculiar
way how conformal symmetry is broken by the introduction of the D7-brane in AdS$_5$ and, 
related to that, the way how this symmetry is preserved in the calculation of special
quantities, like the spin $S$ and the `anomalous dimension' $\gamma$,
which are guaranteed to be scale independent in the dual gauge theory. 
Namely, for a point particle in AdS, the
parameter $mR$ is the product of the two scalars (in the sense of general coordinate 
transformations in AdS$_5$): the `particle mass' $m$ and the `AdS radius' $R$.
By contrast, in the problem of the rotating open string, neither $\omega$ nor $z_0$ are
AdS scalars, rather they separately  change under dilatations, 
but in such a way that their product remains invariant. 

To see this, recall that an infinitesimal dilatation in the boundary gauge theory corresponds to an 
isometry of the bulk theory\footnote{For the explicit form of the AdS isometries in Poincar\' e coordinates, see for example \cite{BoschiFilho:2003zi}.}  with  the following AdS$_5$ transformations:
\begin{eqnarray}\label{dilat}
& &z \to (1+ \epsilon ) z \,\, ;\,\, r \to   (1+ \epsilon ) r \,\, ; \,\, t \to   (1+ \epsilon ) t \,\, ;\cr
& &\,\, \theta \to   \theta \,\, ; \,\,  \omega = \frac{\rmd \theta   }{\rmd t} \to  \omega (1 -\epsilon )  \,\,.
\end{eqnarray}
which leave the tilded variables $ \tilde r$ and  $\tilde z$ unchanged,
This isometry changes the position $z_{_{D7}}$ of the D7-brane according
to $z_{_{D7}}\to (1+ \epsilon ) z_{_{D7}}$, hence it is tantamount to a rescaling
of the mass parameter $M_q$ in the dual gauge theory: $M_q\to   (1 -\epsilon ) M_q$. 
 However, quantities like  $S$ and
$\gamma$, which depend only upon the tilded variables $\tilde z_{_{D7}}$ or  ${\tilde z}_{0}$,
but not on $\omega$ or 
$  z_{D7} \,(= M_q R^2/ 2\pi \alpha' )$ separately,  are invariant under this
transformation, meaning that they are independent of the mass scale $M_q$.
More generally, the string equations of motion and their solutions are invariant under this isometry
when written in terms of the tilded variables. (We implicitly used this symmetry to simplify
the previous analysis.)


This situation should be contrasted with the string calculations of dimension-full 
quantities of the gauge theory, like the size $L$, or of the
spectrum $E$, of the meson states. For a high spin meson, the size is obtained 
as $L\simeq 2C {\tilde z}_0/\omega$. When this result is expressed in terms of the physical quantities
$M_q$ and $S$, one finds $L\sim S^2/(\sqrt{\lambda}M_q)$, which represents a large size, 
$L\gg \sqrt{\lambda}/M_q \sim z_{_{D7}}$, when $S\gg\sqrt{\lambda}$. (Notice that
$\Delta L\sim  z_{_{D7}}$ is the quantum uncertainty in the size of the meson, as predicted by
the UV/IR correspondence in AdS/CFT.) The meson energy is the same as the string energy 
measured by a gauge theory observer (cf. \eqn{EP}), that is \cite{Kruczenski:2003be}
\begin{equation}
\label{boundaryenergy}
E \,= -2\int_{{ z}_{D7}}^{{ z}_0} 
\rmd { z}    \,\pi^0_{0} \,=\,  \frac{R^2 \,\omega }{ \pi \alpha'}\,\,  
 \int_{{\tilde z}_{D7}}^{{\tilde z}_0} 
\rmd {\tilde z} \,  \frac{1 }{{\tilde z}^2} 
\sqrt{\frac{{\tilde r}^{\,\prime \, 2} + 1}{1 - {\tilde r}^{2}}}.
\end{equation}
Clearly, neither $L$ nor $E$ can be expressed as functions of the `tilde' variables alone;
rather, they also depend upon the mass scale $M_q$, as expected for bound states in
a gauge theory. 

One should perhaps emphasize here that the above arguments do not reduce to merely dimensional
analysis, but rather involve subtle aspects of the AdS/CFT dictionary,
like the difference between quantities measured by {\em gauge-theory observers} as opposed to
{\em local observers} in AdS. To see this, notice that if one was to evaluate the string energy
as the sum of the energy densities measured by {\em local observers},  that is (cf. \eqn{warp})
\beq
\bar E \,\equiv -2\int_{{ z}_{D7}}^{{ z}_0} 
\rmd { z}    \,\frac{z}{R}\,\pi^0_{0}\,=\,  \frac{R}{ \pi \alpha'}\,\,  
 \int_{{\tilde z}_{D7}}^{{\tilde z}_0} 
\rmd {\tilde z} \,  \frac{1 }{{\tilde z}} 
\sqrt{\frac{{\tilde r}^{\,\prime \, 2} + 1}{1 - {\tilde r}^{2}}}.
\eeq
then the result $\bar E$ would be independent of the quark mass $M_q$, albeit it would still have
mass dimension one, like the physical energy in \eqn{boundaryenergy}. We have no physical 
interpretation for the dimensionless quantity\footnote{By itself, the `energy' $\bar E$ cannot be
a physical quantity in the gauge theory, since it is explicitly dependent upon the AdS scales
$R$ and $\alpha'$. However, when forming the dimensionless
product $R\bar E$, these dependencies combine into the gauge-theory coupling
factor ${R^2 }{\alpha'}=\sqrt{\lambda}$.} $R\bar E$, but this is 
very similar to our string parameter $\gamma$, as it should be obvious 
from the discussion in Sect.~\ref{sec:gamma}.
 
To summarize, a local measurement  in AdS yields results which 
are independent of the infrared scale and hence can be attributed to a local operator,
whereas a coordinate--time observer measures properties of the gauge-theory state,
which are also sensible to the mass scale $M_q$. So far, we have illustrated these two 
points of view using the examples of the
scale parameter $\gamma$ (`local observer in AdS') and respectively the energy $E$ 
(`gauge theory observer'). It is interesting to similarly consider the case of the spin. 
Indeed, this is at the same time a property of the
physical gauge state, and of the operator creating that state; hence the same value
$S$ must be found both by a gauge theory observer, and by
a local observer in AdS.  This is indeed the case, as we now explain. 

Recall first that, in the case of the energy,
the difference between the string densities measured by the two types of observers is
represented by the warp factor visible in \eqn{warp}. Consider now the respective densities for
the spin. The one seen by the gauge theory observer is
$\rmd S/\rmd z=\pi^0_{\theta} \equiv \partial\cal L/\partial\omega$. As for the local
observer, this measures a spin density
\beq\label{Sbar}
\frac{\rmd\bar S}{\rmd z}\,=\,\frac{\partial\bar{\cal L}}{\partial\bar\omega}\,,\eeq
where $\bar {\cal L} = (z/R)\cal L$ (this follows from $\rmd t \, {\cal L}=
\rmd \bar t\,  \bar{\cal L}$, with $\bar t \equiv \sqrt{g_{00}} \,t= (R/z)t$), while
 \beq
 \bar\omega\,=\,\frac{\rmd\theta}{\rmd \bar t }\,=\,\frac{z}{R}\,\omega\,,\eeq
is the angular velocity measured by the local observer. Hence, \eqn{Sbar} implies
${\rmd\bar S}/{\rmd z}={\rmd S}/{\rmd z}$, and hence $\bar S=S$, as anticipated. 

 \section{Final comments }
 \label{sec:conc}

Through our results in this paper, we brought strong indications in favor of two newly conjectured dualities,
which support each other.  

First, we have proposed a new recipe for computing anomalous
dimensions at infinitely strong coupling for operators, from the ${\cal N} = 4$ or  ${\cal N} = 2$ Yang Mills 
theories, which have large spin or large R-charge, and whose gravity duals are semiclassical strings rotating 
in the AdS$_5\times$S$^5$ space-time (possibly supplemented by D7-branes). Specifically, we 
conjectured a proportionality relation between the anomalous dimensions of interest and the string scale factor 
$\gamma$ introduced in Eq.~(\ref{DimString1}), which is the string proper length in appropriate units.
After testing our recipe against various problems involving closed strings, for which the
anomalous dimensions can be independently and precisely computed via the method pioneered 
in Ref. \cite{Gubser:2002tv}, we have considered the case
of an open string with the endpoints attached to a D7-brane,
for which the traditional method \cite{Gubser:2002tv} fails to apply. 
For this problem, we found that the scale factor $\gamma$ has the same parametric behavior
at large spin as the anomalous dimensions of twist-two operators, as known from closed-string calculations 
(for the single-trace operators built with adjoint fields), and also from the connexion to the cusp anomaly.
This brought us to our second conjectured duality, that between 
a rotating open string with large spin and the quark-antiquark twist-two operator  shown in \eqn{OS}.

An interesting byproduct of our analysis is a distinction between two types of 
physical predictions from open string calculations in AdS$_5$ with D7-branes. 
On one hand, there are observables related to string properties measured by a `boundary' observer 
from the gauge theory; such observables express properties of the dual quantum state, 
like its energy or size, and depend upon the infrared scale in the gauge theory.
On the other hand, there are quantities, like the scale parameter $\gamma$, 
which correspond to local measurements of the string
by an observer living in ten dimensions; these quantities are independent of the mass
scale in the gauge theory and are naturally interpreted as attributes of a local
operator. A quantity like the spin enters both categories (a property of the operator {\em
and} of the quantum state), and indeed it can be computed --- on the string theory
side ---  by either a local, or a boundary, observer, with identical results.

Admittedly, a fundamental understanding of these new dualities is still lacking.
We have no rigorous justification for interpreting $\gamma$  as an anomalous
dimension (for instance, we were not able to associate this quantity with the generator
of some dilatation transformation), nor we understand the limitations of this
interpretation (e.g., we do not know why the proportionality coefficient is not under control).
Also, our current definition of $\gamma$ is restricted the temporal gauge
and it is not clear whether this can be extended to a more general expression, which would
make reparameterization invariance manifest. Also, we do not know 
whether it would make sense to introduce this parameter $\gamma$ for more general strings,
undergoing some complicated motion (e.g. an infinite string with one endpoint attached to an
accelerated heavy quark), and what should be its physical interpretation in that case.
We hope that such questions will trigger further investigations, leading to conceptual
clarifications and to new results.
 
For completeness, let us finally mention that anomalous 
dimensions corresponding to open string configurations with ​large angular momentum $J$, corresponding to​ large ​ $R$-charge, have already been considered 
within the AdS/CFT correspondence, but for a different space--time geometry,
which also involves a plane wave background.
For example,  Ref. \cite{Berenstein:2002zw} discusses open strings in AdS$_5\times$S$^5$/Z$_2$
with angular momentum $J$ in $S^ 5$ space. Both the anomalous dimension and the string length are
found to be independent of $J$,  which is  consistent with our conjecture
of a proportionality relation between these two quantities.  In Ref.   \cite{Stefanski:2003qr}, there is a discussion about open spinning strings with two $S^ 5$  spins in the same, plane wave, background. For these solutions, one can consider different limits of
a large R-charge. We have tested the two particular cases where one keeps one of the spin components fixed, while letting the 
other component to approach infinity. The result is, again, that the anomalous dimension and the string length are independent of the R-charge, in agreement with the conjectured proportionality.


\section*{Acknowledgments} 
We would like to thank Steve Gubser, Juan Maldacena, Horatiu Nastase, and Carlos Nunez for enlightening 
correspondence, and
Al Mueller and Ayan Mukhopadhyay for useful comments on the manuscript.
This work was partially supported by the CAPES--COFECUB project Ph 663/10. 
The work of E.I. is partially supported by the European Research Council under the Advanced 
Investigator Grant ERC-AD-267258.
N.B. would like to thank the Institut de Physique Th\'eorique de Saclay for warm hospitality 
on several occasions, and CNPq for partial support. 
E.I. would like to thank to whole group of the Theoretical Physics Department at UFRJ
(Universidade Federal do Rio de Janeiro) for hospitality at various stages during the completion
of this work.



\begin{thebibliography}{ABC}

\bibitem{Gross:1974cs} 
  D.~J.~Gross and F.~Wilczek,
  Phys.\ Rev.\ D {\bf 9}, 980 (1974).

\bibitem{Georgi:1951sr} 
  H.~Georgi and H.~D.~Politzer,
  Phys.\ Rev.\ D {\bf 9}, 416 (1974).


\bibitem{Korchemsky:1988si} 
  G.~P.~Korchemsky,
  Mod.\ Phys.\ Lett.\ A {\bf 4}, 1257 (1989).

\bibitem{Korchemsky:1992xv} 
  G.~P.~Korchemsky and G.~Marchesini,
  Nucl.\ Phys.\ B {\bf 406}, 225 (1993)
  [hep-ph/9210281].

\bibitem{Basso:2006nk} 
  B.~Basso and G.~P.~Korchemsky,
  Nucl.\ Phys.\ B {\bf 775}, 1 (2007)
  [hep-th/0612247].

\bibitem{Maldacena:1997re}
  J.~M.~Maldacena,
  Adv.\ Theor.\ Math.\ Phys.\  {\bf 2}, 231 (1998)
  [Int.\ J.\ Theor.\ Phys.\  {\bf 38}, 1113 (1999)]
  [arXiv:hep-th/9711200].

\bibitem{Witten:1998qj}
  E.~Witten,
  Adv.\ Theor.\ Math.\ Phys.\  {\bf 2}, 253 (1998)
  [arXiv:hep-th/9802150].

\bibitem{Gubser:1998bc}
  S.~S.~Gubser, I.~R.~Klebanov and A.~M.~Polyakov,
  Phys.\ Lett.\  B {\bf 428}, 105 (1998)
  [arXiv:hep-th/9802109].

\bibitem{Gubser:2002tv} 
  S.~S.~Gubser, I.~R.~Klebanov and A.~M.~Polyakov,
  Nucl.\ Phys.\ B {\bf 636}, 99 (2002)
  [hep-th/0204051].
 
\bibitem{Drukker:1999zq} 
  N.~Drukker, D.~J.~Gross and H.~Ooguri,
  Phys.\ Rev.\ D {\bf 60}, 125006 (1999)
  [hep-th/9904191].

\bibitem{Kruczenski:2002fb} 
  M.~Kruczenski,
  JHEP {\bf 0212}, 024 (2002)
  [hep-th/0210115]. 

\bibitem{Makeenko:2002qe} 
  Y.~Makeenko,
  JHEP {\bf 0301}, 007 (2003)
  [hep-th/0210256].

\bibitem{Belitsky:2003ys} 
  A.~V.~Belitsky, A.~S.~Gorsky and G.~P.~Korchemsky,
  Nucl.\ Phys.\ B {\bf 667}, 3 (2003)
  [hep-th/0304028].

\bibitem{Kruczenski:2007cy} 
  M.~Kruczenski, R.~Roiban, A.~Tirziu and A.~A.~Tseytlin,
  Nucl.\ Phys.\ B {\bf 791}, 93 (2008)
  [arXiv:0707.4254 [hep-th]].

\bibitem{Alday:2007mf} 
  L.~F.~Alday and J.~M.~Maldacena,
  JHEP {\bf 0711}, 019 (2007)
  [arXiv:0708.0672 [hep-th]].



\bibitem{Karch:2002sh} 
  A.~Karch and E.~Katz,
  JHEP {\bf 0206}, 043 (2002)
  [hep-th/0205236].


\bibitem{Kruczenski:2003be} 
  M.~Kruczenski, D.~Mateos, R.~C.~Myers and D.~J.~Winters,
  JHEP {\bf 0307}, 049 (2003)
  [hep-th/0304032].

\bibitem{Erdmenger:2007cm} 
  J.~Erdmenger, N.~Evans, I.~Kirsch and E.~Threlfall,
  Eur.\ Phys.\ J.\ A {\bf 35}, 81 (2008)
  [arXiv:0711.4467 [hep-th]].

\bibitem{Janik:2010gc} 
  R.~A.~Janik, P.~Surowka and A.~Wereszczynski,
  JHEP {\bf 1005}, 030 (2010)
  [arXiv:1002.4613 [hep-th]].

\bibitem{Frolov:2003qc} 
  S.~Frolov and A.~A.~Tseytlin,
  Nucl.\ Phys.\ B {\bf 668}, 77 (2003)
  [hep-th/0304255].

\bibitem{Hofman:2006xt}
D.~M.~Hofman and J.~M.~Maldacena,
J.\ Phys.\ A {\bf 39}, 13095 (2006)
[hep-th/0604135].
 
 
\bibitem{Benna:2006nd} 
  M.~K.~Benna, S.~Benvenuti, I.~R.~Klebanov and A.~Scardicchio,
  Phys.\ Rev.\ Lett.\  {\bf 98}, 131603 (2007)
  [hep-th/0611135].

 
\bibitem{Maldacena:1998im} 
  J.~M.~Maldacena,
  Phys.\ Rev.\ Lett.\  {\bf 80}, 4859 (1998)
  [hep-th/9803002].
   
\bibitem{Rey:1998ik} 
  S.~-J.~Rey and J.~-T.~Yee,
  Eur.\ Phys.\ J.\ C {\bf 22}, 379 (2001)
  [hep-th/9803001].
  
\bibitem{BoschiFilho:2003zi} 
  H.~Boschi-Filho and N.~R.~F.~Braga,
  Class.\ Quant.\ Grav.\  {\bf 21}, 2427 (2004)
  [hep-th/0311012].
  
  
\bibitem{Berenstein:2002zw} 
  D.~E.~Berenstein, E.~Gava, J.~M.~Maldacena, K.~S.~Narain and H.~S.~Nastase,
  hep-th/0203249.
 
\bibitem{Stefanski:2003qr} 
  B.~Stefanski, Jr.,
  JHEP {\bf 0403}, 057 (2004)
  [hep-th/0312091].
  
 
\end{thebibliography}
 \end{document}